 \documentclass[pmlr,twocolumn,10pt]{jmlr} 

\usepackage{booktabs}
\usepackage{siunitx}
\usepackage{algorithm2e} 
\newtheorem*{hyp*}{Hypothesis \protect\hypnumber} 
\newenvironment{hyp}[1]{\renewcommand{\hypnumber}{#1}\begin{hyp*}}{\end{hyp*}}
\newcommand{\hypnumber}{}
\usepackage[switch]{lineno}

\theorembodyfont{\upshape}
\theoremheaderfont{\scshape}
\theorempostheader{:}
\theoremsep{\newline}

\jmlrvolume{LEAVE UNSET}
\jmlryear{2024}
\jmlrsubmitted{LEAVE UNSET}
\jmlrpublished{LEAVE UNSET}
\jmlrworkshop{Preprint} 

 \title[Auditing Language Models in Healthcare]{A Clinical Trial Design Approach to Auditing Language Models in Healthcare Setting }

 \author{%
  \Name{Lovedeep Gondara} \Email{lovedeep.gondara@phsa.ca}\\
  \Name{Jonathan Simkin} \Email{jonathan.simkin@bccancer.bc.ca}\\
  \Name{Shebnum Devji} \Email{shebnum.devji@phsa.ca}\\
 }

\begin{document}

\maketitle

\begin{abstract}
We present an audit mechanism for language models, with a focus on models deployed in the healthcare setting. Our proposed mechanism takes inspiration from clinical trial design where we posit the language model audit as a single blind equivalence trial, with the comparison of interest being the subject matter experts. We show that using our proposed method, we can follow principled sample size and power calculations, leading to the requirement of sampling minimum number of records while maintaining the audit integrity and statistical soundness. Finally, we provide a real-world example of the audit used in a production environment in a large-scale public health network.
\end{abstract}

\paragraph*{Data and Code Availability}
This paper uses identifiable real patient data which is not made publicly available. Any researchers interested in obtaining the data are encouraged to contact authors. As the paper introduces an audit mechanism, there is no code. Access to the models audited should be requested via email to the authors.

\paragraph*{Institutional Review Board (IRB)}
This study was approved by the Research Ethics Board (REB) at University of British Columbia - BC Cancer Research Ethics Board (REB) (REB number H23-01445).

\section{Introduction}
The rapid advancements in machine learning, especially natural language processing, has led to the development of sophisticated language models such as BERT \citep{devlin2018bert}, Longformer \citep{beltagy2020longformer}, Llama \citep{touvron2023llama}, etc. These models have demonstrated exceptional capabilities to address a diverse array of tasks, from classification to natural language generation. This success has motivated researchers to investigate the use of language models in healthcare \citep{thirunavukarasu2023large}, where models have been developed to facilitate clinical documentation \citep{mulyar2019phenotyping}, classifying electronic health record documents \citep{mascio2020comparative}, predicting patient’s survival \citep{nunez2023predicting}, etc.

Language models, similar to any other machine learning model, are prone to mistakes \citep{zhang2023language}. Hallucinations in generative language models \citep{zhang2023siren} have garnered significant attention, where the generative models can generate wrong or invalid text not grounded in the training data. However, erroneous classifications carry the same risk. Classification errors made by the language models in critical domains such as healthcare can be related to an erroneous diagnosis, treatment, or any other model output. Due to the nature of the language models, these errors are often produced with high confidence \citep{zhao2021calibrate}. The errors can occur due to various underlying reasons, such as an ill-informed training process, a biased training dataset, etc. Regardless of the reason, classification errors pose a significant risk of providing wrong information, which can have serious downstream consequence in the healthcare domain.

A principled method to keep a consistent check on language models is auditing. Model auditing can be described as a mechanism that developers and policy makers use for risk management and mitigation associated with the deployment of language models \citep{mokander2023auditing}. Model auditing for language models is at its infancy \citep{mokander2023auditing}, where the proposed approaches mostly focus on generative models for toxicity and bias \citep{bender2021dangers}. There is a paucity of literature when it comes to evaluating language models deployed in healthcare, especially in the case of classification and information extraction.

We, at the population-based British Columbia Cancer Registry (BCCR) within a large public healthcare system have traditionally used subject matter experts (SMEs) to review pathology reports for classifying them as reportable tumors (included in the cancer registry database) vs. non-reportable tumors (excluded from the cancer registry database). As of 2020, we have transitioned to the use of eMaRC software (Electronic Mapping, Reporting, and Coding) \citep{emarcplus}, a rule-based text processing framework developed by the US Centers for Disease Control and Prevention, which is specialized for identifying digital pathology reports indicative of a reportable tumour.

The eMaRC software, however, suffers from the drawbacks of a rule-based system \citep{gehrmann2017comparing}, where it fails to capture the intricacies of complex pathology reports, resulting in classifying significant number of non-reportable tumors as reportable. This results in extensive manual review by the SMEs, leading to delays in information capture, impeding research, opportunities to support patient care and quality improvement initiatives, and timely health system operations that rely on up-to-date data. In an effort to improve the timeliness of the cancer registry database, and  address increasing volumes of source data and significant resource constraints, the BCCR transitioned to the use of language models to identify pathology reports indicative of a potential reportable tumor from a real-time population-based electronic feed of pathology reports.  

In our case, the decisions made by the language models directly impact disease surveillance, cancer hospital operations, and research. We need to ensure that the language models deployed in production are working as intended and to the standards set by the national and international bodies. At BCCR, we have developed a novel model auditing process which involves manual review of the pathology reports by the SMEs. Where SMEs are unaware if the report has been previously classified as a reportable tumor by either the language models or by eMaRC. SMEs, instead, follow the standard reportability guidelines set by the national and international bodies. 

Human evaluation of language models is not new \citep{chang2024survey,bang2023multitask}. Our innovation lies in the audit design of the language models, where our process mimics a single blind clinical trial. The SMEs are unaware of the previously assigned reportability status to minimize bias. The audit process is setup as an equivalence trial, with the null hypothesis that the language models and SMEs have a statistically significantly different performance, with the performance margin of at least $\delta$. This allows us to utilize principled sample size and power calculations to ensure that the sample size for audit is as small as possible while providing us with statistically sound conclusions, making the audit operationally feasible.

The main message in this paper is two-fold, first it showcases the significance of principled model audit approaches in critical domains such as the healthcare, and second, it shows the indispensable importance of SMEs. That is, the importance of SME-in-the-loop model auditing, where SMEs are required to ensure the deployed models are working as intended. Lastly, we demonstrate practical application of our audit mechanism when applied to the production environment of a large publicly funded health system. We show how the proposed audit helps to identify system discrepancies, giving us warning to intervene at earlier stage, ensuring continued high standard of patient care and data collection.

\section{Why Audit and Related Work}
Language models have revolutionized various aspects of healthcare delivery \citep{thirunavukarasu2023large}, offering unprecedented opportunities for improving patient outcomes, optimizing resource allocation, and enhancing clinical decision-making. However, the integration of these models into healthcare workflows introduces unique challenges related to their accuracy, interpretability, and ethical implications. To address these challenges, auditing procedures have emerged as indispensable mechanisms for evaluating the performance, fairness, and compliance of language models. However, the research  still is in its infancy \citep{mokander2023auditing}. Next, we briefly discuss the need for auditing language models in healthcare production environments and provide some the general related work. 

\begin{itemize}
    \item \textbf{Accuracy and Reliability.} One of the primary motivations for auditing language models in the healthcare setting is to ensure their performance and reliability in clinical decision-making. Errors or biases within these models can have profound consequences on patient care, potentially leading to misdiagnosis, inappropriate treatment selection, or adverse outcomes. In the general setting, \citet{shi2023retrieval} proposes human-in-the-loop evaluation for system validation and improvement, emphasizing the importance of evaluating language models in diverse contexts. \citet{jones2023automatically} proposes automated auditing of language models via discrete optimization where the model jointly optimizes over inputs and outputs and helps to uncover various failure modes.

    \item \textbf{Transparency and Interpretability.} The opacity of deep learning, often referred to as the ``black box" problem, poses significant challenges to the adoption and acceptance of language models within the healthcare setting. Auditing procedures play a pivotal role in promoting transparency and interpretability by elucidating the underlying mechanisms driving the model's predictions. \citet{ribeiro2016model} developed LIME (Local Interpretable Model-agnostic Explanations), a framework for generating explanations for model predictions at the instance level. \citet{lundberg2017unified} introduced SHAP (SHapley Additive exPlanations), a method for explaining individual predictions using game theory principles.

    \item \textbf{Bias Detection and Mitigation.}  The propensity for language models to perpetuate biases present in the underlying data represents a critical concern in healthcare, where equitable treatment and fair outcomes are paramount. Auditing methodologies enable the detection and mitigation of biases by examining the composition and distribution of training data, evaluating model performance across diverse demographic groups, and implementing corrective measures to address disparities. \citet{bolukbasi2016man} introduced methods for quantifying and mitigating gender biases in word embeddings, highlighting the importance of addressing biases embedded in the training data. \citet{dixon2018measuring} proposed metrics for quantifying biases in language models, facilitating the identification of biased language generation patterns.

    \item \textbf{Compliance and Regulation.}  Healthcare organizations operate within a complex regulatory landscape characterized by stringent privacy requirements, ethical standards, and quality assurance measures. Auditing ML models ensures compliance with regulatory frameworks such as the Health Insurance Portability and Accountability Act (HIPAA) \citep{hipaa}, the General Data Protection Regulation (GDPR) \citep{GDPR2016a}, and ethical guidelines outlined by professional associations and governing bodies. By verifying adherence to legal and ethical principles, auditing safeguards patient privacy, confidentiality, and rights within the context of machine learning-driven healthcare systems. In the general setting, several studies have explored the ethical and societal compliance and implications of language models. \citet{bender2021dangers} highlighted the societal biases present in large language models and advocated for responsible AI practices to mitigate these biases. \citet{mitchell2019model} emphasized the importance of considering the broader societal impacts of language technologies, including issues related to fairness, privacy, and accessibility.
\end{itemize}

\section{The Audit}
When the language models are ready to be deployed, one of the critical challenges is assessing the performance of the model in real-world scenarios. That is \emph{our focus} in this paper. Despite rigorous testing during development, the true efficacy of a model can only be validated through post-deployment monitoring. Upon deployment, the operational environment introduces nuances that may affect the model's performance unpredictably. This post-deployment validation is especially important in critical domains such as the healthcare where the model decisions directly impact patient care.

To bridge this gap between anticipated and actual performance, manual verification becomes imperative. In domains such as healthcare, SMEs are required to scrutinize the model outputs against real-world observations using their expertise, which enables them to discern discrepancies, thereby assessing the model's accuracy and reliability. The use of SMEs is necessary because a lay person is not qualified to understand clinical text and make the appropriate decision (diagnosis, prognosis, etc.). SMEs undergo years of training to develop these skills. The manual assessment by the SMEs lays the foundation for our audit process.

The main idea behind the use of SMEs for the audit process is that when SMEs evaluate the model's performance, we can compare the model outputs with expert knowledge, which serves as the gold-standard. Using this approach we can determine the correctness of the model's predictions. The efficacy of this approach is contingent upon the following questions.

\begin{enumerate}
    \item How many data points are needed for the audit process?
    \item How do we know with confidence that the model is performing as expected?
\end{enumerate}

To answer these questions, we propose a principled audit design following the principles of clinical trial design. In the following section, we introduce our general audit approach, followed by an example of implementing the proposed audit in a large scale publicly funded healthcare system.

\subsection{Audit Design}
Our proposed audit design takes inspiration from the clinical trial literature, where we posit our  audit design in-spirit similar to a single blind clinical trial. Where in our case, blinding is done at the SME level where the SMEs do not know if the clinical document they are viewing has been classified by the language model beforehand. This blinding is done to minimize any potential bias where the SME’s decision can be influenced by the prior decision made by the language models. 

To begin the audit process, we take a random sample of the documents that are to be classified (further details in Section \ref{sec:aud_design}). This random sample of documents is then sent to SMEs for manual review and to the language models for classification. The decisions from SMEs and the models are stored in a secure database for further analysis.

This design, in principle, replicates a scenario where a single entity, a clinical document in our case, undergoes multiple interventions, and we are interested in comparing the efficacy of one intervention vs. the other. The interventions in our case being being classified by the language model and by the SMEs. In a clinical trial setting, the entity is a person, and if a single person is receiving multiple interventions (treatments) that we want to compare, we have to account for the confounding due to the impact of one treatment on the other.  

In our case, the entity is a document, and hence we do not have that issue. As the document being classified by the language model does not impact the decision made by the SME for the same document (because they are blinded to the decision made by the language model), this scenario is akin to a randomized clinical trial where patients are randomized into multiple treatment arms, and the confounding due to underlying factors is accounted for by the use of randomization. In our case, the SMEs and the language models reviewing the same document ensure that the impact of confounding variables in our case (document length, language complexity, etc.) is the same for all.

\begin{figure*}[t]
    \centering
    \includegraphics[scale=0.6]{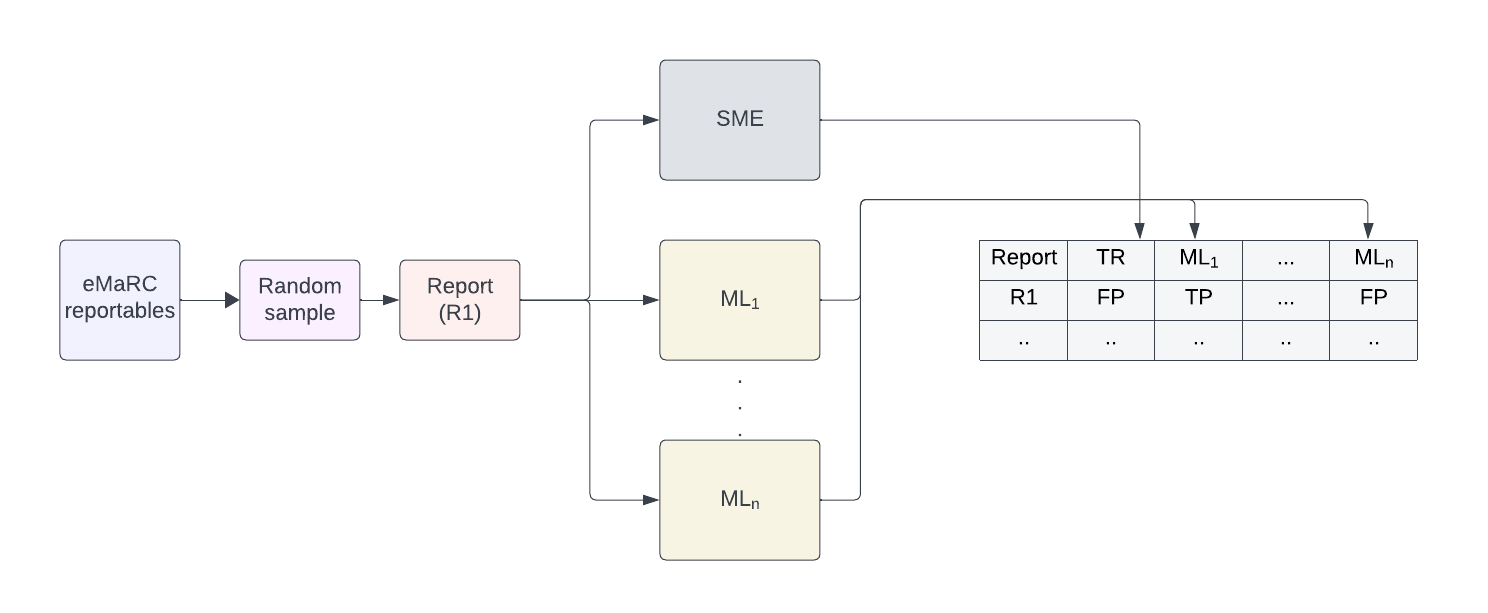}
    \caption{The audit design. A random sample of reports classified as reportable by the eMaRC system are reviewed by the SMEs and are classified by the language models. The results are stored in a database for further analysis.}
    \label{fig:auditdesign}
\end{figure*}

\subsection{The Hypothesis}\label{sec:hyp}
Our main comparison of interest is SMEs vs the language model. There are multiple clinical trial designs to choose from, we select the equivalence trial design due to its many inherent advantages, some of which we detail below.

\begin{itemize}
    \item \textbf{Direct assessment of similarity.} Equivalence trial will help us directly assess whether the model performance is not significantly worse than the SMEs by establishing equivalence margins. This provides a clear indication of whether the models are comparable to the SMEs, without claiming superiority. This is relevant to our use-case where our aim is to show that the models are just as effective as SMEs. Our goal is not to show that the models can outperform SMEs, our goal is to have an automated method that performs at-par within a predefined margin, and can be scaled up to handle the volume of data encountered in a large healthcare system.

    \item \textbf{Reduced sample size requirements.} Equivalence trials typically require smaller sample sizes compared to superiority trials, as they focus on detecting differences within a predefined margin rather than detecting \emph{any} difference. This is arguably the most important property for a model audit because the audit requires manual input by SMEs whose time is a finite resource. 

    \item \textbf{Minimized bias in interpretation.} Superiority trials can suffer from biases in interpretation, where the investigators focus on statistically significant differences even if they are not practically meaningful (Example: A difference of 0.5\% accuracy might be statistically significant, but not practically useful.). Equivalence trials, on the other hand, provide a clearer framework for interpreting results within the context of predefined equivalence margins, minimizing bias.
\end{itemize}
With the above in mind, we setup our hypotheses as follows.
\begin{hyp}{(Null)}
SMEs and language models have statistically significantly different performance, with the performance difference of at least $\delta$.
\end{hyp}

\begin{hyp}{(Alternate)}
SMEs and language models do not have statistically detectable difference of at least $\delta$.  
\end{hyp}

\subsection{Sample Size Calculations}\label{sec:samp_size}
The next step in the trial design is the sample size calculations. Sample size calculations are a critical component of the trial design, ensuring that the study is adequately powered to detect meaningful differences between the models' performance and the SMEs, if any exists. The calculation of sample size in equivalence trials involves considerations of statistical power, significance level, expected variability, and the chosen margin of equivalence ($\delta$ in hypotheses above). We use the following formula for the sample size estimation \citep{zhong2009calculate}.

\begin{equation}
    N = 2 \times (\frac{z_{1 - \alpha/2}+z_{1 - \beta}}{\delta})^2 \times p \times (1 - p)
\end{equation}

Where $\delta$ is our equivalence margin, $p$ is the perceived accuracy of SMEs, and $z$ is the standard normal deviate. $\alpha$ and $\beta$ are the standard parameters in statistical sample size calculations and correspond to the elements of Type I and Type II errors.

\section{Audit in Practice}\label{sec:aud_design}
To demonstrate the proposed audit process at work in practice at the BCCR, we use the original problem discussed in the Introduction. That is, classifying pathology reports as reportable tumors vs. non-reportable tumors.

Figure \ref{fig:auditdesign} shows the conceptual design of this audit process. We take a random sample of the reports classified as the reportable tumor by the rule-based eMaRC system. This random sample of reports is then sent to SMEs for manual review and to the language models for classification. This process is shown in Figure \ref{fig:auditdesign} using a single report R1 and the language models (ML$_1$,$\cdots$, ML$_n$). SMEs and the language models make a decision on R1 independently whether it is a reportable tumor (true positve, TP) or a non-reportable tumor (false positive, FP).

\subsection{Our Hypothesis}
Based on the information provided in Section \ref{sec:hyp}, our null and alternate hypothesis are as follows.
\begin{hyp}{(Null)}
For reportability classification, SMEs and language models have statistically significantly different classification performance, with the performance difference of at least $\delta$.
\end{hyp}

\begin{hyp}{(Alternate)}
SMEs and language models do not have statistically detectable difference of at least $\delta$ for reportability classification.  
\end{hyp}

Using the sample size calculations described Section \ref{sec:samp_size}, we use $\delta=0.02$, that is, a performance difference within $2\%$ of the SMEs and the models would be acceptable to us. For $p$, we assume it is $98\%$, which is based on our earlier estimates using SMEs. Assuming an $\alpha=0.05$ and a $\beta=0.025$, i.e. a power of $97.5\%$, we get a sample size of $1506$. That is, we need to audit $1506$ pathology reports to reach a meaningful conclusion.

\subsection{Data Background and Initial Model}
This section provides the required background on the BCCR, the BCCR dataset, the status-quo of rule-based systems, the requirement for a deep learning based approach, and the round one of deep learning based model training and deployment.

\subsubsection{British Columbia Cancer Registry (BCCR)}
The BCCR collects a population-based dataset containing personal and tumour information on all reportable tumours among the residents of BC. The registry conforms to national and international standards for the classification of cancer records. 

As of the year 2020, the BCCR receives all provincial pathology reports as Health Level Seven (HL7, an industry standard) messages in real-time from regional laboratory information systems, including cancer and non-cancer reports. The BCCR applies exclusions to incoming HL7 messages. Exclusions include out of province residents, invalid messages, and cancer screening related messages on pre-malignant lesions. The remaining HL7 messages are sent for processing to eMaRC (detailed next). 

\subsubsection{eMaRC}\label{sec:emarc}
Electronic Mapping, Reporting, and Coding (eMaRC) \citep{emarcplus} is a software developed by the Centers for Disease Control and Prevention. BCCR uses eMaRC (version 6.0.0.5) to distinguish between reportable tumours, non-reportable tumours, and non-cancer pathology. The eMaRC is a rule-based text analysis and classification platform. Rule-based approaches are not robust and often underperform compared to deep learning approaches \citep{gehrmann2017comparing}. eMaRC is no exception, where it results in a large volume of false positives. That is, eMaRC classifies many tumor pathology reports as reportable cancer when the reports are truly non-reportable. 

SMEs at the BCCR manually review all reports that eMaRC considers reportable (approximately 127,000 per year) to identify false positives (approximately 38,000 per year). This is a monumental task involving significant operational resources. This drawback of eMaRC lead to the development and deployment of deep learning based models to automate the classification of tumor pathology reports.

\subsubsection{Initial Training Dataset}\label{sec:dataset1}
The initial training dataset for the language model to classify reportable vs. non-reportable tumors consists of 40,000 eMaRC classified reportable pathology reports from 2021. The reports were manually reviewed to assign gold-standard labels as to if the reports were a true-reportable tumor or a false positive. Special attention was paid to sample the false positive reports to capture the special categories of non-reportability such as the non-reportable skin cancers. Random partitions of the dataset were used as the validation (10\%) and test set (20\%) to evaluate the performance of the model.

\subsubsection{Initial Model}
Our initial model used a multi-tiered approach, where after preprocessing, relevant segments of the pathology report were extracted and used as an input to a BERT-based \citep{Vaswani+2017}  Natural Language Inference (NLI) model, where the output from NLI were the various features (non-reportable histologies, non-reportable sites, reportable histologies, reportable sites, etc.) that make a pathology report a reportable tumor vs non-reportable. These features were then used to train a random forest classifier for the final task of binary prediction (true reportable vs. false positive). Probability thresholding based on a hold-out validation set was used in the random forest classifier to boost the true-reportable performance to acceptable levels. 

\subsubsection{Initial Model Results}
Evaluating the model on the test partition showed great results with 98\% per-class accuracy on the true-reportable tumors and 90\% per-class accuracy on the false positives. These accuracy results satisfy the operational benchmark set by the BCCR and are supportive of the idea to deploy the models in the production environment.

\subsection{Administering Audit}
After successfully testing the model on the hold-out test set, the next step is to use the proposed audit mechanism to measure the model's performance compared to SMEs. One of the main challenges is administering the audit, keeping in the mind that the SMEs have limited time and are non-technical. We facilitated the audit process using a Graphical User Interface (GUI), offering a user-friendly platform for seamless interaction. Notably, this interface maintained confidentiality by concealing whether the reports under evaluation had been previously assessed by either the ML models or the eMarC system. 

Within this controlled environment, SMEs were allotted a predetermined time frame (1 minute) to review each pathology report, during which they classified pathology reports as either reportable or non-reportable. The outcomes of these evaluations were securely stored in a designated database.  

\subsection{Initial Audit Results}
The audit showed our deployed model was underperforming compared to SMEs. Despite rigorous testing and validation during development, the difference between the expected and actual performance in a real-world setting was significant. Based on the model's performance on the test set, we expected the model to maintain its test accuracy of 98\% on the reportable tumors.

However, the audit of our production environment showed accuracy of 85\% for reportable tumor classification. Accuracy for classifying the non-reportable tumors decreased to 70\%. Based on our hypothesis, the results were well outside the margin ($\delta$)\footnote{Evident by the confidence interval between the difference of performance of SMEs and the model being outside $(-\delta, \delta$), i.e. ($-2\%, 2\%$)}, leading us to fail to reject our null hypothesis, and reaching the conclusion that the language model has statistically significantly different performance compared to SMEs, with the performance of the language model being lower. This failure indicated a critical flaw, in either the model development or deployment procedures. 

\subsection{The Reason for Failure}
After a thorough investigation, we found that there were multiple reasons for the model's diminished performance. We provide details on them next.

\begin{itemize}
    \item \textbf{Reason 1 (Biased training data).} The first reason was the method used for curating initial training dataset (Section \ref{sec:dataset1}). The data used to train the model had unintentional built-in biases. When creating the training dataset, we tried to cover all possible types of non-reportable tumors. This lead to oversampling of some classes of non-reportable tumors that are otherwise underrepresented in the real-world. As a result, the model struggled to generalize to the real-world situations, even after performing well on the test data. This shows the importance of having good quality, representative, and diverse datasets when training language models. 

    \item \textbf{Reason 2 (Flawed model design).} The second reason for failure was the flawed model development process where the initial model was unnecessarily complex using the combination of NLI and a random forest with random forest using hardcoded probability thresholding based on the validation dataset to reach a certain accuracy. This resulted in the thresholds being invalid in the real-world scenario. The complexity of the modelling process also led to obscuring the inner workings of the model, making debugging harder. This taught us an important lesson to keep models simple when possible. 
\end{itemize}

\section{Intervention}
The results from the audit process prompted us to begin work on interventions, where the first step was to curate a representative dataset.

\begin{itemize}
    \item \textbf{Intervention 1 (The New Training Data).}  As the first intervention to address the shortcomings of the initial dataset, we created a new training dataset comprised of 40,000 pathology reports from the year 2021, selected using a random sampling method. In contrast to cherry picking the non-reportable cases, this approach ensured the dataset represents true distribution with respect to the real-world. Mirroring the composition of the complete dataset, the training data had a distribution of 65\% true reportable tumors and 35\% false positives. 

    \item \textbf{Intervention 2 (The New Model).} Based on lessons learned, we deployed a simpler, ensemble approach using two BERT-based models. Where the models are finetuned on the new training data described above and combined using an OR-based approach. This approach is used to boost true reportable accuracy and is shown in Table \ref{tab:combine_logic} in the Appendix \ref{apd:first}.
\end{itemize}

We would like to emphasize that as the focus of this paper is the proposed auditing approach, we have not provided fine-grained details on the models that are being audited. Our proposed approach is model-agnostic and would work the same with any architecture.

\subsection{Evaluating the Intervention}
For the initial evaluation of the models using the test data, we selected random sample of pathology reports from multiple years (2021, 2022, and 2023). Using a total of 30,400 reports, we observe an average 98\% accuracy on reportable tumors and 92\% accuracy on non-reportable tumors.

\subsection{Auditing the Intervention}
In order to validate the performance of our newly deployed models within real-world context, we need to repeat the audit process. The new audit yielded promising outcomes, where we observed that the models are functioning as expected, mirroring the accuracy we observed on the test sets in the real-world. These results led to the rejection of the null hypothesis, with the evidence in favor of the alternate hypothesis, that is, the SMEs and the deployed model do not have statistically detectable difference of $\delta$\footnote{Evident by the confidence interval between the difference of performance of SMEs and the model being in between $(-\delta, \delta$), i.e. ($-2\%, 2\%$)}. These results reinforce the efficacy of our current deployed models. These models are presently operational at the BCCR, efficiently processing and classifying tens of thousands of pathology reports in real time, contributing to streamlined automated data collection and decision-making processes.

\section{Making Audits Operationally Feasible}
Executing a large-scale audit, where a large sample size is required for statistical soundness, presents significant operational challenges. The challenges primarily stem from operational constraints that impede the simultaneous allocation of SME resources. The main reasons being the limited supply of SMEs and the operational costs associated with utilization of SME time for audit vs other clinical duties. To circumvent this issue and to ensure the continuous and comprehensive monitoring of model performance, we have adopted a piecemeal audit approach. Where the audit process has been segmented into manageable units, each representing a fraction of the total sample size, thereby facilitating ongoing audits conducted incrementally over time, only using a small portion of the SMEs operational time. 

Specifically, the total sample size has been partitioned into twelve equal chunks, with SMEs tasked with auditing one chunk per month. This systematic division enables the prompt indication of any deviations from the anticipated model performance, thereby facilitating early and frequent interventions. Moreover, by the year's end, the accumulation of data from all audit chunks enables robust statistical analysis based on the total sample size calculations. 

Notably, this methodology mirrors the principles of Bayesian clinical trial designs, wherein patient accrual occurs continuously, allowing for iterative analyses and adaptations throughout the trial period. Such iterative auditing not only ensures the timely indication of any deviations and anomalies, but also fosters a dynamic and responsive approach to maintaining model efficacy and reliability in real-world applications.

\section{Conclusion}
This paper has introduced a novel audit methodology for language models within the healthcare domain inspired by the principles from clinical trial design, specifically adopting a framework akin to a single-blind clinical trial. By leveraging existing literature for sample size calculations and power analysis, our proposed audit design minimizes the number of samples required for manual auditing, a resource-intensive process reliant on SMEs. Through empirical validation using a real-world scenario, we have demonstrated the efficacy of our proposed method in practice. 

\bibliography{jmlr-sample}

\newpage
\appendix
\section{Appendix}\label{apd:first}
\subsection{The Model Combination Approach}
In this approach, a report is classified as a reportable tumor if either of the models predict it as a reportable tumor. To inject variance into the model output and to ensure that both models cover as much relevant text as possible within the pathology report, both models use slightly different data pre-processing methods. Where one model focuses on particular segments within the report and the other model simply uses the top 512 tokens (the token limit for BERT-based models). 
\begin{table}[h]
    \centering
    \begin{tabular}{c|c|c}
        Model 1 & Model 2 & Combined \\ \hline
        Reportable & Reportable & Reportable\\
        Non-Reportable & Reportable & Reportable \\
        Reportable & Non-Reportable  & Reportable \\
        Non-Reportable & Non-Reportable & Non-Reportable \\
    \end{tabular}
    \caption{Combining the output from both models to maximize true reportable accuracy.}
    \label{tab:combine_logic}
\end{table}

\end{document}